\documentclass[a4paper]{jpconf-kb}
\usepackage{amsmath}
\usepackage{iopams}
\usepackage{natbib}
\usepackage{graphicx}

\usepackage{multind}
\makeindex{subject}
\makeindex{authors}

\newcommand{\bfrt}{({\bf{r}},t)}

\begin{document}

\title{Quantum turbulence in atomic Bose-Einstein condensates}

\author{A. J. Allen, N. G. Parker,  N. P. Proukakis, and C. F. Barenghi.}

\address{Joint Quantum Centre (JQC) Durham-Newcastle, School of Mathematics and Statistics,\newline
\phantom{$^1$ }Newcastle University, Newcastle upon Tyne NE1 7RU, England, UK.}

\ead{joy.allen@ncl.ac.uk, nick.parker@ncl.ac.uk, nikolaos.proukakis@ncl.ac.uk, carlo.barenghi@ncl.ac.uk}

\index{authors}{Allen, A. J.} \index{authors}{Parker, N. G.} \index{authors}{Proukakis, N. P.}
\index{authors}{Barenghi, C. F.}

\begin{abstract}
%
      
      
 Weakly interacting, dilute atomic Bose-Einstein condensates (BECs) have proved to be an attractive context
for the study of  nonlinear dynamics and 
quantum effects at the macroscopic scale.  Recently, weakly interacting, dilute atomic BECs have been used 
to investigate quantum
turbulence\index{subject}{quantum turbulence!turbulence} both experimentally 
and theoretically, stimulated largely by the high degree of control which
is available within these quantum gases.  
In this article we motivate the use of weakly interacting, dilute atomic BECs for the study of turbulence, 
discuss the characteristic regimes of turbulence which are accessible,  
and briefly review some selected investigations of quantum turbulence and
recent results.  We focus on three stages of turbulence - the generation of 
turbulence, its steady state and its decay - and highlight some fundamental 
questions regarding our understanding in each of these regimes.  

\end{abstract}

%


\section{Introduction}
      Turbulence in ordinary fluids such as air or water consists of rotational eddies of different sizes
      which we term vortices\index{subject}{vortices}.  Vortices\index{subject}{vortices!quantum vortices} therefore
      are a hallmark signature of a turbulent flow~\citep{Barenghi01}.\index{authors}{Barenghi, C. F.}\index{Donnelly,
      R. J.}     
      In superfluids, quantum
      vortices\index{subject}{vortices!quantum vortices} differ from their classical counterparts because of the
      quantization of circulation.   This means that
      the rotational motion of a superfluid is constrained to discrete
      vortices\index{subject}{vortices!quantum vortices} which all have the same core structure.  Turbulence in
      superfluid helium has been the subject of many recent
      experimental and theoretical investigations~\citep{Skrbek12}\index {authors}{Sreenivasan, K. R.}. 
      Recently, experimentalists have been
      able to visualise individual vortex lines and reconnection events
      using tracer particles~\citep{Fonda12}\index{authors}{Fonda, E.}\index{authors}{Meichle,
      D.P.}\index{authors}{Ouellette, N. T.}\index{authors}{Hormoz, S.}\index{authors}{Sreenivasan, K. R.}.  Weakly
      interacting, dilute atomic Bose-Einstein
      condensates (henceforth referred to as BECs)\index{subject}{Bose Einstein condensates} present a distinct platform to view and probe
      quantum turbulence\index{subject}{quantum turbulence!turbulence}.  A key feature here is the ability to
      directly resolve the structure of individual vortices\index{subject}{vortices!quantum vortices} and in
      turn the dynamics of a turbulent vortex tangle~\citep{Henn09}.\index{authors}{Henn, E. A. L.}
      \index{authors}{Seman, J. A.}\index{authors}{Roati, G.}\index{authors}{Magalh\~aes, K. M. F.}
      \index{authors}{Bagnato, V. S.}\index{subject}{vortices!vortex tangle}
      As a result of the quantized nature of vorticity, 
quantum turbulence in superfluid helium 
and in BECs\index{subject}{Bose-Einstein condensates} can be viewed as a 
      simpler, idealized analog of turbulence in ordinary fluids, 
and opens the possibility of studying problems
      which may be relevant to our general understanding of turbulence.

\section{Why atomic Bose-Einstein condensates?}
     Since their first generation 
     in 1995~\citep{Davis95,Anderson95},\index{authors}{Davis, K. B.}\index{authors}{Mewes, M. O.}\index{authors}{Andrews, M. R.}\index{authors}{van Druten, N. J.}\index{authors}{Durfee, D. S.} 
\index{authors}{Kurn, D. M.}\index{authors}{Ketterle, W.}\index{authors}{Anderson, M. H.}
\index{authors}{Ensher, J.}\index{authors}{Matthews, M. R.}\index{authors}{Wieman, C. E.} atomic BECs have been used to study a wide variety of
     nonlinear dynamics, for example, solitons, vortices\index{subject}{vortices!quantum vortices} and four-wave
     mixing  \citep{Kevrekidis08}\index{authors}{Kevrekidis, P. G.}\index{authors}{Carretero-Gonzalez, R.}.  
A merit of exploiting BECs as a testbed of nonlinear physics lies with 
the immense control and flexibility they offer. For example: 
     \begin{itemize}
       \item \emph{Trapping geometry, shape and dimensionality}\\
       Optical and magnetic fields can be employed to precisely create a potential landscape for the atoms in the BEC,
       which in turn enables control of the shape and effective dimensionality of the system
       \citep{Gorlitz01}.\index{authors}{G\"orlitz, A.}\index{authors}{Vogels, J. M.}\index{authors}{Leanhardt, A. E.}
       \index{authors}{Raman, C.}\index{authors}{Gustavson, T. L.}\index{authors}{Abo-Shaeer, J. R.}
      \index{authors}{Chikkatur, A. P.}\index{authors}{Gupta, S.}\index{authors}{Inouye, S.}\index{authors}{Rosenband, T.}
      \index{authors}{Ketterle, W.}  A
       basic requirement of these gases is confinement in space to prevent contact with hot surfaces.  This is
       typically provided by magnetic traps which are harmonic in shape and have the form
       \citep{Fortagh07}\index{authors}{Fort\'agh, J.}\index{authors}{Zimmermann, C.}
     \begin{eqnarray}
       V_{\rm{ext}}({\bf r}) = \frac{1}{2}m\omega (x^2 + y^2 + z^2),
     \end{eqnarray}
     where $\omega$ is the trapping frequency and $m$ the mass of the atom.
     This type of trap results in an atomic cloud with a radial density profile which resembles an inverted parabola.

     If a harmonic trap is used which is very strongly confining in one direction, for example
      \begin{eqnarray}
       V_{\rm{ext}}({\bf r}) = \frac{1}{2}m\omega (x^2 + y^2 + \epsilon z^2),
     \end{eqnarray}
     where $\epsilon \gg 1$, the dynamics in that direction
     is effectively inhibited and the system becomes effectively two-dimensional (2D).  By changing
     $\epsilon$, one can easily change the effective dimensionality, which is particularly important
     in turbulence (2D turbulence is very different from 3D turbulence).  
     In the same way, if the trap is 
     very tightly confining in two directions, the dynamics 
     is mainly in the third direction, and the system is effectively 
one-dimensional (1D).  

    More complicated trapping geometries can be realised, for example 
     a toroidal ring or a periodic optical lattice. Traps can also
be made time-dependent by rotating or shaking the trapping potential.  
Furthermore, one can create localized potentials using optical fields, which can mimic an obstacle and be moved through the system on demand.  

   \item \emph{Interaction strength}\\
   Typically, the dominant atomic interaction in a BEC\index{subject}{Bose-Einstein condensates} is the short-range
   and isotropic {\it s}-wave interaction.  Experimentalists can employ magnetic Feshbach
   resonances~\citep{Inouye98}\index{authors}{Inouye, S.}\index{authors}{Andrews, M. R.}\index{authors}{Stenger,
   J.}\index{authors}{Miesner, H. J.}\index{authors}{Stamper-Kurn, D. M.}\index{authors}{Ketterle, W.} to change the strength of
   these interactions and even their nature,  i.e. whether they are attractive or repulsive
   \citep{Roberts98}\index{authors}{Roberts, J. L.}\index{authors}{Claussen, N. R.}\index{authors}{Burke, J.
   P.}\index{authors}{Greene, C. H.}\index{authors}{Cornell, E. A.}.  Furthermore, by using atoms with relatively large
   magnetic dipole moments, e.g. $^{52}$Cr, it is possible to create a BEC\index{subject}{Bose-Einstein condensates}
   where the atoms also experience significant dipole-dipole interactions, which are long-range and anisotropic, and
   greatly modify the static and dynamical properties of the system \citep{Lahaye09}\index{authors}{Lahaye, T.}\index{authors}{Menotti, C.}\index{authors}{Santos, L.}\index{authors}{Lewenstein, M.}\index{authors}{Pfau, T}.
     
   \item \emph{Vortex core optical imaging} \\
     The healing length which characterizes the vortex core size is typically around $10^{-7}$m in a BEC\index{subject}{Bose-Einstein condensates} (c.f.
     $10^{-10}$m in superfluid Helium).  By expanding the BEC\index{subject}{Bose-Einstein condensates} (following
   release from the trapping potential), the vortex can be directly imaged and resolved via optical
 absorption~\citep{Madison00,Raman01}\index{authors}{Madison, K. W.}\index{authors}{Chevy, F.}\index{authors}{Wohlleben, W.}\index{authors}{Dalibard, J.}\index{authors}{Raman, C.}\index{authors}{Abo-Shaeer, J. R.}\index{authors}{Vogels, J. M.}\index{authors}{Xu, K.}\index{authors}{Ketterle, W.}.  Advanced real time imaging of condensates
     containing vortices\index{subject}{vortices!quantum vortices} has also recently been developed~\citep{Freilich10}\index{authors}{Freilich, D. V.}\index{authors}{Bianchi, D. M.}\index{authors}{Kaufman, A. M.}\index{authors}{Langin, T. K.}\index{authors}{Hall, D. S.}, allowing the precession of vortices\index{subject}{vortices!quantum vortices} to be observed.

   \end{itemize}

     In the limit of zero temperature and weak interactions, the evolution equation for the macroscopic 
     condensate wavefunction, $\phi \bfrt$, is a form of nonlinear Schr{\"o}dinger equation, commonly known
     as the Gross-Pitaevskii Equation (GPE)\index{subject}{Gross-Pitaevskii Equation}:
     \begin{eqnarray}
       i\hbar \frac{ \partial \phi(\mathbf{r},t)}{ \partial t} = \left[ - \frac{\hbar ^2}{2m} \nabla^2 +
       V_{\rm ext}(\mathbf {r},t)  + g|\phi(\mathbf{r},t)|^2 - \mu \right] \phi(\mathbf{r},t),
     \label{eqn_gpe}
     \end{eqnarray}
     where $\bf{r}$ is the position in space, $t$ is time and $\hbar$ is Planck's constant divided by $2\pi$.
     The GPE\index{subject}{Gross-Pitaevskii Equation} provides a good quantitative description of the dynamics of a BEC\index{subject}{Bose-Einstein condensates} over all lengthscales
     available, from the vortex core to the system size, up to temperatures of approximately 
     $T \simeq 0.5T_c$ (where the critical temperature $T_c$ is of the order of mK in typical
     experiments).   At the right-hand side we recognise the kinetic energy term, $(-\hbar^2/2m) \nabla^2$, 
     the trapping potential $V_{\rm ext}(\mathbf {r},t)$ (which in general may be time-dependent), the interaction term,  $g|\phi(\mathbf{r},t)|^2$ where
     $g = 4\pi \hbar^2 a_s/m$ and $a_s$ is the 3D $s-$wave scattering length, and the chemical potential $\mu$.
     The GPE\index{subject}{Gross-Pitaevskii Equation} can be almost exactly mapped to the classical Euler equation; 
the small difference, namely the
     quantum stress, regularizes 
     the solutions, preventing singularities which may arise in an Euler
     fluid~\citep{Barenghi08}\index{authors}{Barenghi, C. F.}.
The GPE\index{subject}{Gross-Pitaevskii Equation} is a practically exact model in the limit of zero temperature, where
essentially all of the atoms exist in the Bose-Einstein condensate phase\index{subject}{Bose-Einstein condensates}. In many experiments the condensate exists at well below the BEC\index{subject}{Bose-Einstein condensates} transition temperature such that this approximation is justified.
     Extensions of the GPE\index{subject}{Gross-Pitaevskii Equation} to include the 
     effect of thermal atoms provide a more complete (albeit not exact) physical model of a real
     BEC~\citep{Jackson09}\index{authors}{Jackson, B.}\index{authors}{Proukakis, N. P.}
     \index{authors}{Barenghi, C. F.}\index{authors}{Zaremba, E.} (see
     e.g.~\cite{Proukakis08}\index{authors}{Proukakis, N. P.}\index{authors}{Jackson, B.} for an in depth review of finite
     temperature models).
     

     However BECs\index{subject}{Bose-Einstein condensates} suffer an
important limitation.  The systems which can be currently created in the 
laboratory contain a small number of atoms, 
     typically $10^3$ to $10^9$, hence do not sustain the number of 
quantum vortices\index{subject}{vortices!quantum vortices} present in helium 
experiments.  For example, up to a few hundred vortices\index{subject}{vortices!quantum vortices} have been achieved in the largest 2D BECs \citep{Abo-shaeer02}\index{authors}{Abo-Shaeer, J. R.}\index{authors}{Raman, C.}\index{authors}{Ketterle, W.}.
     This brings to light the issue of length scales.  A defining
     feature of classical turbulence, besides nonlinearity, is the huge number of length scales which are excited.  The range of length scales available in 
turbulent superfluid helium at very small temperatures is perhaps
even larger, since short wavelength helical waves 
     along the vortices\index{subject}{vortices!quantum vortices} can be
generated by nonlinear interactions, producing a turbulent cascade
called the Kelvin wave cascade~\citep{Vinen06}\index{authors}{Vinen, W.}.  A simple question arises: 
\emph{can a BEC, containing a limited number of 
vortices\index{subject}{vortices!quantum vortices}, really become turbulent?} 
Our tentative answer is yes.  
Numerical results thus far \citep{Nore97,Berloff02,Kobayashi05b,Yepez09}\index{authors}{Nore, C.}\index{authors}{Abid, M.}\index{authors}{Brachet, M. E.}\index{authors}{Berloff, N. G.}\index{authors}{Svistunov, B. V.}\index{authors}{Kobayashi, M.}\index{authors}{Tsubota, M.}\index{authors}{Yepez, J.}\index{authors}{Vahala, G.}\index{authors}{Vahala, L.}\index{authors}{Soe, M.} 
suggest that kinetic energy is distributed over the length scales in agreement
with the $k^{-5/3}$ Kolmogorov scaling which is observed in ordinary turbulence
(where $k$ is the wavenumber) even over this small range of length scales. 
     Therefore, the study of turbulence in a BEC\index{subject}{Bose-Einstein condensates} represents an exciting
     opportunity to probe a new regime residing somewhere between chaos and turbulence.   
     
     In the remainder of this paper, we aim to identify some important open questions about
     turbulence in BECs; where appropriate we will
    review some of the work which has been carried out to date.

    \section{Quantum Turbulence in atomic BECs, where are we?}
    \label{sec:qt_BEC}
    The following is an extensive, but by no means exhaustive, 
    list of aspects yet to be understood regarding turbulence 
in atomic BECs\index{subject}{Bose-Einstein condensates}.  
    To structure our discussion, we distinguish
    the evolution of turbulent flow into three stages, namely;
    \begin{itemize}
      \item[(i)] \underline{The generation of the turbulence.}   \emph{What are the most effective and efficient ways to generate turbulence?} \emph{Does the way in which the turbulence is generated affect the `type' of turbulence created?} 
      \item[(ii)] \underline{The statistical steady state.} \emph{Are there universal features of turbulence, for instance, is the Kolmogorov energy spectrum present?  What are the statistics of the turbulence velocity field?}
    \item[(iii)] \underline{The decay of the turbulence.}  \emph{How does the turbulence decay?} \emph{What is the best way to measure the decay?}
    \end{itemize}~\\
       

    \noindent{\bf{(i)  The generation of the turbulence}}\\
    To understand the generation of a vortex 
tangle\index{subject}{vortices!vortex tangle} in a quantum gas, we must 
first understand how individual 
vortices\index{subject}{vortices!quantum vortices} are nucleated.  
The very first creation of such a vortex took place in
    a two-component condensate and was driven by the rotation of one component around 
    the other.  The subsequent removal of the inner component resulted in the formation of a hollow core of a 
    singly quantized vortex~\citep{Anderson00}\index{authors}{Anderson, B. P.}\index{authors}{Haljan, P. C.}\index{authors}{Wieman, C. E.}\index{authors}{Cornell, E. A.}.  Further techniques 
    for generating vortex structures soon followed, including the creation of vortex rings following the ``snake instability" decay of a dark soliton ~\citep{Anderson01}\index{authors}{Anderson, B. P.}\index{authors}{Haljan, P. C.}\index{authors}{Regal, C. A.}\index{authors}{Feder, D. L.}\index{authors}{Collins, L. A.}\index{authors}{Clark, C. W.}\index{authors}{Cornell, E. A.}, 
    phase imprinting~\citep{Leanhardt02}\index{authors}{Leanhardt, A. E.}\index{authors}{G\"orlitz, A.}\index{authors}{Chikkatur, A. P.}\index{authors}{Kielpinski, D.}\index{authors}{Shin, Y.}\index{authors}{Pritchard, D. E.}\index{authors}{Ketterle, W.} and by
    a rapid quench through the transition temperature for the onset of
    Bose-Einstein condensation\index{subject}{Bose-Einstein condensates}, i.e. the Kibble-Zurek mechanism~\citep{Weiler08,Freilich10}\index{authors}{Weiler, C.}\index{authors}{Neely, T. W.}\index{authors}{Scherer, D. R.}\index{authors}{Bradley, A. S.}\index{authors}{Davis, M. J.}\index{authors}{Anderson, B. P.}.

    However, to generate a large number of vortices\index{subject}{vortices!quantum vortices} in the system at any one time, 
    two other techniques have proved to be more effective:
    \begin{itemize}
      \item[(i)] \emph{Rotation} of an anisotropic BEC\index{subject}{Bose-Einstein condensates} excites surface modes 
      leading to the nucleation of vortices at the edge which then drift into the bulk of the BEC.  If the rotation is performed about only one axis, a vortex lattice\index{subject}{vortices!lattice} is created
       ~\citep{Hodby01,Abo-shaeer02,Abo-shaeer01,Madison00,Madison01}\index{authors}{Hodby, E.}\index{authors}{Hechenblaikner,
       G.}\index{authors}{Hopkins, S. A.}\index{authors}{Marag\`o, O. M.}\index{authors}{Foot, C. J.}\index{authors}{Abo-Shaeer, J. R.}\index{authors}{Raman, C.}\index{authors}{Ketterle, W.}\index{authors}{Abo-Shaeer, J. R.}\index{authors}{Raman, C.}\index{authors}{Vogels, J. M.}\index{authors}{Ketterle, W.}\index{authors}{Madison, K.
W.}\index{authors}{Chevy, F.}\index{authors}{Wohlleben, W.}\index{authors}{Dalibard, J.}\index{authors}{Madison, K.
W.}\index{authors}{Chevy, F.}\index{authors}{Bretin, F.}\index{authors}{Dalibard, J.}. 
In 3D, if the rotation is performed about more than one axis,
       a vortex tangle\index{subject}{vortices!vortex tangle} has
       been predicted to form~\citep{Kobayashi07}\index{authors}{Kobayashi, M.}\index{authors}{Tsubota, M.}.  
      \item[(ii)] \emph{A moving (cylindrical) obstacle}, such as that created by the potential from a blue-detuned laser beam moving 
      through a quantum fluid, generates
      pairs of vortices\index{subject}{vortices!quantum vortices} 
in its wake when its speed exceeds a critical value~\citep{Raman01}\index{authors}{Raman, C.}\index{authors}{Abo-Shaeer, J. R.}\index{authors}{Vogels, J. M.}\index{authors}{Xu, K.}\index{authors}{Ketterle, W.}.  Recently, this method has been used to generate and study a collection of vortex dipoles in a 2D BEC~\citep{Neely10}\index{authors}{Neely, T. W.}\index{authors}{Samson, E. C.}\index{authors}{Bradley, A. S.}\index{authors}{Davis, M. J.}\index{authors}{Anderson, B. P.}.  

    \end{itemize}

\begin{figure}[h!]
      \centering{
          \includegraphics[scale = 0.5]{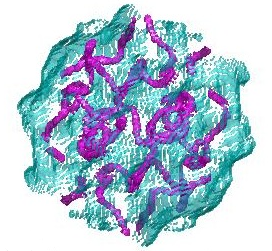}
       }
       \caption{3D turbulent state of a harmonically trapped BEC.  Condensate edge shown by blue shading, turbulent
       vortex tangle\index{subject}{vortices!vortex tangle} shown in purple \citep{White10}\index{authors}{White, A.
       C.}\index{authors}{Barenghi, C. F.}\index{authors}{Proukakis, N. P.}\index{authors}{Youd, A.
       J.}\index{authors}{Wacks, D. H.}. Figure courtesy of Angela C. White.  
     }
       \label{white_tangle}
     \end{figure}

    Both methods generate a large number of 
vortices\index{subject}{vortices!quantum vortices}, in 2D as well as in 3D.  However, one can 
    bypass the initial transient period and begin with a nonequilibrium state of vortices\index{subject}{vortices!quantum vortices}.  Experimentally, this can be achieved via 
    imprinting a phase profile onto the condensate via laser beams, as performed by~\cite{Leanhardt02} for
    generating single vortices\index{subject}{vortices!quantum vortices} of arbitrary charge.  The use of such imprinting to generate a vortex tangle has
    been implemented theoretically~\citep{White10}, with the resulting tangle similar to that depicted in Fig~\ref{white_tangle}.
    \\

    Our preliminary results with method (ii) (laser stirring), suggest that it is possible to
    generate a large number of vortices\index{subject}{vortices!quantum vortices}; we have found qualitative evidence
    that, by moving the obstacle along different paths, 
    we can change the isotropy of the resulting tangle of vortices.  
    
    \begin{figure}[h!]
    \centering{
     a)
    \includegraphics[scale = 0.3]{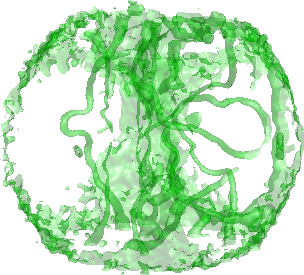}
    \hspace{0.1cm}
    b)
    \includegraphics[scale = 0.3]{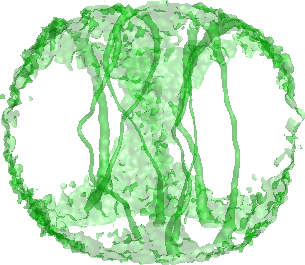}
    \hspace{0.1cm}
    c)
    \includegraphics[scale = 0.3]{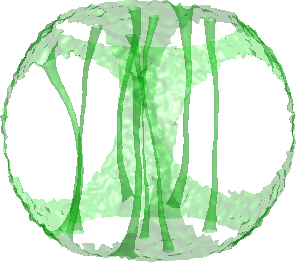}
    }\\
    \includegraphics[clip,scale = 0.4,angle = 270]{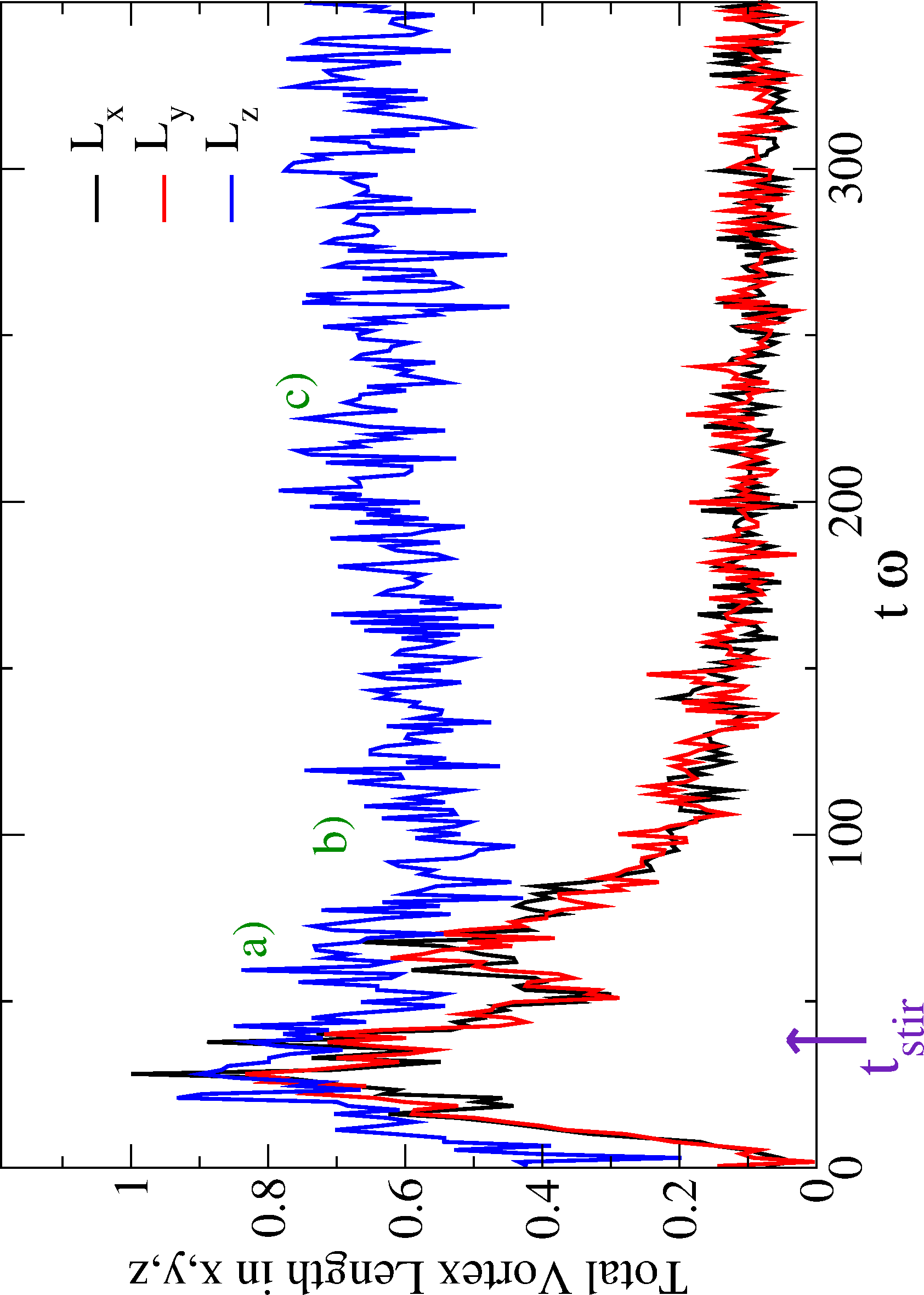}
    \caption{Top row:  Density isosurfaces of a 3D spherical BEC\index{subject}{Bose-Einstein condensates} at times
      $t\omega = $ a) 60, b) 100 and c) 240 after stirring
    the condensate along the $z-$direction in a circular path.  We see here that the surface plot picks up the vortex
    cores as well as some of the condensate edge.  The resulting vortex length in each direction is shown over time
    in the bottom part of this figure where it has been normalised by the peak vortex length.}
    \label{fig:iso_cir}
     \end{figure}
    Fig.~\ref{fig:iso_cir} (top row) shows the density isosurface of a 3D spherical condensate at three 
    instants in time, after the condensate has been stirred for a time $t_{\rm{stir}}$, along a
    circular path with a Gaussian-shaped laser stirrer oriented in the $z-$direction.  
For a simple measure of the isotropy of the 
tangle\index{subject}{vortices!vortex tangle}, we plot the projected vortex lengths
$L_x$, $L_y$ and $L_z$ in each Cartesian direction 
(bottom part of Fig.~\ref{fig:iso_cir}).  
All projected lengths rapidly increase during the stirring period 
($t< t_{\rm{stir}}$); after the laser has been removed ($t> t_{\rm{stirr}}$),
$L_x$, $L_y$ and $L_z$ all decrease for a short period of time.
Later, only the vortex lengths $L_x$ and $L_y$ in the transverse 
($x$ and $y$) directions further decrease, whereas $L_z$ remains
approximately constant because the vortex tangle decays into a 
regular lattice, as it is apparent from the final density isosurface 
(Top row, c)).   This is as expected: in stirring the condensate circularly we impart angular
momentum about the $z$-axis, and it is well known that the ground state of a superfluid with
sufficient angular momentum features a lattice of regularly-spaced, vortex lines aligned along the
$z$-axis. 
    
    \begin{figure}[h!]
      \centering{
         \includegraphics[scale = 0.4,angle = 270]{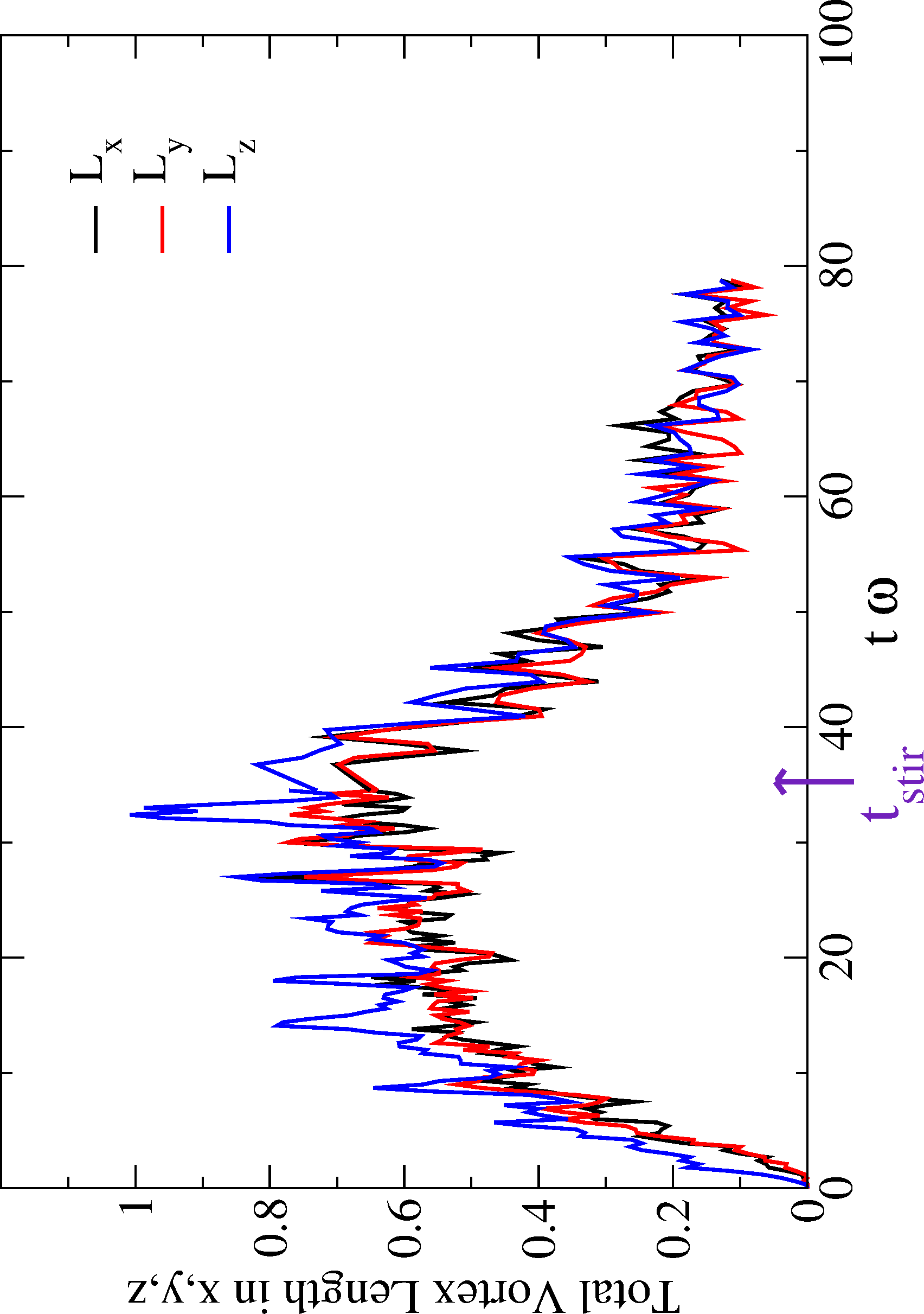}
            }
            \caption{Vortex length, normalised by the peak length, in the $x,y$ and $z$ directions for a stirrer
            along the $z-$direction, moving in a Figure-eight path.}
            \label{fig:infin}
    \end{figure}

    In Fig.~\ref{fig:infin} the vortex length is shown when the stirring 
takes place, for the same amount of time, along a Figure-eight path.  
Again, the vortex length increases over the duration of the laser stirring 
($t< t_{\rm{stir}}$); however, after the laser is removed ($t> t_{\rm{stir}}$),
the tangle decays isotropically, i.e. all projected lengths $L_x$,
$L_y$ and $L_z$ decay together.
For the Figure-eight path, the laser also moves through the centre 
of the condensate, where the density is higher, the vortices which
are generated are longer than those generated  
at the edge of the condensate; therefore, this laser stirring path 
is more efficient at creating a dense, random vortex 
tangle\index{subject}{vortices!vortex tangle} than simply stirring the
    condensate in a circular fashion. \\~\\

    \noindent{\bf{(ii) Statistical state}}\\
    Once the turbulence state has been created (by whichever means), 
its steady state properties can be investigated.
BECs offer the possibility to study 2D and 3D turbulence and the cross-over 
region between the two \citep{Parker05}\index{authors}{Parker, N. G.}\index{authors}{Adams, C. S.}.  We now review our current understanding of the properties
    of a turbulent tangle of vortices\index{subject}{vortices!quantum vortices}\index{subject}{vortices!vortex tangle} in a BEC, first in 2D and then in 3D. \\

    \noindent\emph{Quantum turbulence in 2D}\\
    In 2D classical turbulence, the conservation of enstrophy dictates that energy must flow 
    from small scales of energy injection to large scales forming, for example, large 
    clusters of vortices~\citep{Kraichnan67}\index{authors}{Kraichnan, R. H.}\index{subject}{vortices!quantum vortices}.  This inverse cascade is 
    thought to underly Jupiter's giant
    Red Spot and has been experimentally examined in classical 
    fluids~\citep{Sommeria88,Marcus88}\index{authors}{Sommeria, J.}\index{authors}{Meyers, S. D.}\index{authors}{Swinney, H. L.}\index{authors}{Marcus, P.} (for a review see~\cite{Kellay02}\index{authors}{Kellay, H.}\index{authors}{Goldburg, W. I.}).  Attempts to observe the inverse cascade
    effect in quantum gases  have lead to
    modelling vortex generation~\citep{Parker05,White12,Fujimoto11,Reeves13},\index{authors}{Parker, N. G.}
   \index{authors}{Adams, C. S.}\index{authors}{White, A. C.}\index{authors}{Barenghi, C. F.}
    \index{authors}{Proukakis, N. P.}\index{authors}{Fujimoto, K.}\index{authors}{Tsubota, M.}\index{authors}{Reeves, M.}
    \index{authors}{Billam, T.}\index{authors}{Anderson, B. P.}\index{authors}{Bradley, A. S.} and to experiments on
    the dynamics of vortex dipoles created by a moving
potential~\citep{Neely10,Neely12}.\index{authors}{Neely, T. W.}\index{authors}{Samson, E.
C.}\index{authors}{Bradley, A. S.}\index{authors}{Davis, M. J.}\index{authors}{Anderson, B.
P.}\index{authors}{Neely, T. W.}\index{authors}{Bradley, A. S.}\index{authors}{Samson, E. C.}
\index{authors}{Rooney, S.}\index{authors}{Wright, E.}\index{authors}{Law,
K.}\index{authors}{Carretero-Gonzalez, R.}\index{authors}{Kevrekidis, P. G.}\index{authors}{Davis,
M. J.}\index{authors}{Anderson, B. P.} 

    ~\cite{Numasato10}\index{authors}{Numasato, R.}\index{authors}{Tsubota, M.}\index{authors}{L'vov, V. S.}, evolved the 2D GPE\index{subject}{Gross-Pitaevskii Equation} to a turbulent state by initially 
    imposing a random phase on the wavefunction.  They did not observe a reverse 
    cascade but rather a direct cascade.  They argued that since the
    total number of vortices\index{subject}{vortices!quantum vortices}, and therefore the enstrophy, is not conserved in simulations of the GPE\index{subject}{Gross-Pitaevskii Equation} because
    of vortex pair annhilation, the inverse energy cascade is irrelevant for 2D quantum
    turbulence\index{subject}{quantum turbulence!turbulence}.  

    Conversely, ~\cite{Reeves13}\index{authors}{Reeves, M.}\index{authors}{Billam, T.}\index{authors}{Anderson, B. P.}\index{authors}{Bradley, A.
S.} reported the numerical observation of the inverse cascade.  
    They solved the 2D damped GPE\index{subject}{Gross-Pitaevskii Equation} (DGPE) and generated a turbulent state 
    by imposing the fluid to flow past 4 stationary potential barriers.  The speed of the flow was 
    sufficiently high ($v \simeq 0.822c$, where $c$ is the sound speed of the quantum fluid) so as to create
    many vortices\index{subject}{vortices!quantum vortices} and thereby a turbulent flow.
    Using a statistical algorithm they measured how prone `like-winding' vortices\index{subject}{vortices!quantum vortices}, i.e. 
    vortices\index{subject}{vortices!quantum vortices} of the same charge, were to cluster together depending on the amount of 
    dissipation imposed.  They found that an intermediate level of damping lead 
    to small clusters of like-winding vortices\index{subject}{vortices!quantum vortices} being formed and inferred an inverse energy
    cascade from analysis of the incompressible energy spectrum (energy associated with vortices\index{subject}{vortices!quantum vortices}).  

Similar work by~\cite{White12}\index{authors}{White, A. C.}\index{authors}{Barenghi, C. F.}\index{authors}{Proukakis, N. P.} implemented a rotating elliptical paddle 
to generate large numbers of 
vortices\index{subject}{vortices!quantum vortices}.  
Again, clustering of like-winding 
vortices\index{subject}{vortices!quantum vortices} was observed, but no
inverse cascade was reported.

A possible reason for the lack of definitive evidence for or against
the inverse cascade in quantum gases is their relatively small size, 
i.e. the systems which were studied lacked enough length scales.  
However, this limitation has not prevented the observation of the direct 
Kolmogorov-3D cascade in systems of similar 
size~\citep{Nore97,Kobayashi05a,Kobayashi05b,Yepez09,Kobayashi07}\index{authors}{Nore, C.}\index{authors}{Abid,
M.}\index{authors}{Brachet, M. E.}\index{authors}{Kobayashi, M.}\index{authors}{Tsubota, M.}\index{authors}{Kobayashi,
M.}\index{authors}{Tsubota, M.}\index{authors}{Yepez, J.}\index{authors}{Vahala, G.}\index{authors}{Vahala,
L.}\index{authors}{Soe, M.}\index{authors}{Kobayashi, M.}\index{authors}{Tsubota, M.}.\\
       

\noindent\emph{Quantum turbulence in 3D}\\
In their seminal work, ~\cite{Nore97}\index{authors}{Abid, M.}\index{authors}{Brachet, M. E.} used the GPE\index{subject}{Gross-Pitaevskii Equation} to investigate 3D 
quantum turbulence\index{subject}{quantum turbulence!turbulence} 
in a homogeneous box by evolving an initial, large scale Taylor Green vortex.
By decomposing the velocity field into divergence-free and 
curl-free parts, they obtained incompressible 
(associated with vortex motion) and compressible 
(associated with acoustic excitations) kinetic energy spectra respectively.
They showed that the incompressible kinetic
energy spectrum  is similar to 
the classical $k^{-5/3}$ Kolmogorov energy spectrum at 
scales down to the intervortex spacing.  
    Similarly,~\cite{Berloff02}\index{authors}{Berloff, N. G.}\index{authors}{Svistunov, B. V.} began with a non-equilibrium state, generated 
    by imprinting a random phase on the equilibrium wavefunction of a homogeneous box.  
    They then observed the evolution of the quantum turbulence\index{subject}{quantum turbulence!turbulence} by solving the GPE\index{subject}{Gross-Pitaevskii Equation} and further 
    allowing the system to evolve to phase coherence.  
    Similarly,~\cite{Kobayashi05a,Kobayashi05b}\index{authors}{Kobayashi, M.}\index{authors}{Tsubota, M.}\index{authors}{Kobayashi, M.}\index{authors}{Tsubota, M.} imprinted a random phase on
    the equilibrium wavefunction of a homogeneous box and evolved according to the GPE\index{subject}{Gross-Pitaevskii Equation}.  
    They introduced a dissipative term
    which only acted on scales smaller than the healing length 
    to represent thermal dissipation in the system. They also obtained a decaying incompressible energy 
    spectrum which has the Kolmogorov power law over the inertial range.  
In order to clarify the extent of this range, statistical steady turbulence 
was created by a moving random potential which continuously injected
energy into the system at large scales; a damping term removed energy
at small length scales.  They found that 
the inertial range was slightly narrower for the continuously forced
    turbulence because the moving potential sets the 
    energy containing range.  
    Further to these methods, \cite{White10}\index{authors}{White, A. C.}\index{authors}{Barenghi, C. F.}\index{authors}{Proukakis, N. P.}\index{authors}{Youd, A. J.}\index{authors}{Wacks, D. H.} imprinted a 
    staggered vortex array onto a
    harmonically trapped BEC\index{subject}{Bose-Einstein condensates} and evolved the system according to 
    the GPE\index{subject}{Gross-Pitaevskii Equation} (see Fig.~\ref{white_tangle}).  
In this work, they calculated the probability density function of the velocity
components and found that (in both 2D and 3D) it is not a Gaussian like
in ordinary turbulence, in agreement with experimental results obtained 
in superfluid helium~\citep{Paoletti08}\index{authors}{Paoletti, M. S.}\index{authors}{Fisher, M. E.}\index{authors}{Sreenivasan, K. R.}\index{authors}{Lathrop, D. P.}.
    This is an important observation which distinguishes quantum from classical turbulence.
    
    Using a similar method of imprinting,~\citep{Yepez09}\index{authors}{Yepez, J.}\index{authors}{Vahala, G.}\index{authors}{Vahala, L.}\index{authors}{Soe, M.} performed 
    impressively large simulations in a homogeneous box using a 
quantum lattice gas algorithm (up to $5760^3$ grid points) and resolved
scales smaller than the vortex core radius.
   
    Most of the methods of generating quantum 
turbulence\index{subject}{quantum turbulence!turbulence} discussed so far 
in this section have a common aim: the incompressible
energy spectrum of quantum turbulence\index{subject}{quantum turbulence!turbulence} after an initial, turbulent state has been set up 
    (with the exception of~\citep{Kobayashi05a,Kobayashi05b}\index{authors}{Kobayashi, M.}\index{authors}{Tsubota, M.}\index{authors}{Kobayashi, M.}\index{authors}{Tsubota, M.}).  
    However, the group of Tsubota have also carried out
    simulations where they dynamically create a vortex tangle\index{subject}{vortices!vortex tangle} by solving the DGPE\index{subject}{Gross-Pitaevskii Equation} with combined rotation
    along two axis of a harmonically trapped BEC. They  found that by changing the ratio between the rotation
    frequencies in both directions, they could generate a 
    vortex lattice\index{subject}{vortices!lattice} or a more disordered array of
    vortices\index{subject}{vortices!quantum vortices} which formed a vortex tangle\index{subject}{vortices!vortex tangle} in
    which individual 
    vortices\index{subject}{vortices!quantum vortices} appear to be nucleated with no preferred direction.
    They measured the incompressible energy spectrum and found it to be
    consistent with Kolmogorov law.

Experimentally,  a small vortex 
tangle\index{subject}{vortices!vortex tangle} has been created 
in a harmonically trapped BEC\index{subject}{Bose-Einstein condensates} 
through the combination of rotation and an external oscillating 
perturbation
by Bagnato's group~\citep{Henn09,Henn10,Seman11,Shiozaki11}.\index{authors}{Henn, E. A.
L.}\index{authors}{Seman, J. A.}\index{authors}{Roati, G.}\index{authors}{Magalh\~aes, K. M.
F.}\index{authors}{Bagnato, V. S.}\index{authors}{Henn, E. A. L.}\index{authors}{Seman, J. A.}
\index{authors}{Roati, G.}\index{authors}{Magalh\~aes, K. M. F.}\index{authors}{Bagnato, V. S.}
They noticed that upon expansion of the condensate
the usual inversion of aspect ratio of the gas~\citep{Mewes96,Castin96} did not happen.
\index{authors}{Mewes, M. O.}\index{authors}{Andrews, M. R.}\index{authors}{van Druten, N. J.}
\index{authors}{Kurn, D. M.}\index{authors}{Durfee, D. S.}\index{authors}{Ketterle, W.}
\index{authors}{Castin, Y.}\index{authors}{Dum, R.} This effect
could be a possible signature of the creation of a tangle of 
vortices\index{subject}{vortices!vortex tangle}.  

    This brings us to the final stage of the evolution of turbulence, the decay.\\~\\

    \noindent{\bf{(iii) The decay of the turbulence}}\\
 In classical turbulence, the cascade of kinetic energy over the length scales 
terminates at some very short scale where viscosity dissipates kinetic energy
into heat.  The absence of viscosity in quantum fluids
means there must exist other mechanisms of energy dissipation. The most
likely is acoustic emission.
When two vortices\index{subject}{vortices!quantum vortices} reconnect, 
some energy is lost in the form of sound.  Reconnections 
are also thought to create high frequency Kelvin waves on 
vortices\index{subject}{vortices!quantum vortices}.  It is thought that,
in superfluid helium,
Kelvin waves interact nonlinearly and create shorter and shorter waves,
until sound waves are emitted at high frequency~\citep{Vinen06}\index{authors}{Vinen, W.}. This
energy transfer is called the Kelvin wave cascade.

Experiments in superfluid helium show that, depending on the scale at which 
energy is injected, the decay of the turbulence can be one of two 
forms~\citep{Baggaley12}\index{authors}{Baggaley, A. W.}\index{authors}{Barenghi, C. F.}\index{authors}{Sergeev, Y. A.}:
    \begin{itemize}
      \item[(i)] \emph{`Semiclassical' or `Kolmogorov' turbulence:}  The vortex tangle\index{subject}{vortices!vortex tangle}
 seems polarised and 
    structured over many length scales.  This type of turbulence is 
generated when the forcing is at length scales larger than the average
intervortex spacing.  In this regime, the vortex length $L$ decays
as $L \simeq t^{-3/2}$, which is consistent with the decay of a Kolmogorov
spectrum.
  \item[(ii)] \emph{`Ultraquantum' or `Vinen' turbulence:} The scale of
        the forcing is less than the intervortex spacing,
    the vortex tangle\index{subject}{vortices!vortex tangle} seems random and possesses a single length scale, and the vortex 
    length decays as $L \simeq t^{-1}$~\citep{Walmsley08}\index{authors}{Walmsley, P. M.}\index{authors}{Golov, A. I.}. 
    \end{itemize}

    How to measure the decay in a BEC is an open question.  
    \cite{White10}\index{authors}{White, A. C.}\index{authors}{Barenghi, C. F.}\index{authors}{Proukakis, N. P.} 
    \index{authors}{Youd, A. J.}\index{authors}{Wacks, D. H.} studied the decay of the turbulent
    tangle\index{subject}{vortices!vortex tangle} by numerically monitoring the vortex length, $L$, over time.  
    They showed that the line length increases initially 
    as reconnections take place before decaying 
    over time.  By further solving the dissipative GPE\index{subject}{Gross-Pitaevskii Equation}, they confirmed that 
    thermal dissipation leads to a faster decay of line length but could not clearly distinguish between
    $L \simeq t^{-1}$ or $L \simeq e ^{-\alpha t}$ behaviour ($\alpha$ being some decay parameter).
    However, is vortex linelength the best measure for this decay? Or can we again look to the 
    incompressible energy spectrum to visualise the decay of vortices\index{subject}{vortices!quantum vortices} and draw some conclusions
    from both of these quantities?  Furthermore, there is the question of how this is best achieved
    experimentally.  The images taken of condensate density are typically column-integrated over
    the imaging direction which means that depth information becomes lost and an extraction of the
    true 3D vortex line length is not possible.  Just as the attenuation of second sound is used to
    measure vortex length in helium, what surrogate measures of vortex line length are accessible
    experimentally in BECs\index{subject}{Bose-Einstein condensates}?  Such questions, both
    fundamental and practical in nature, will provide a rich source of research in these systems in the future.

  \section{Summary}
 We have discussed weakly interacting, dilute atomic Bose-Einstein condensates\index{subject}{Bose-Einstein condensates} as tools for understanding the nature of
 quantum turbulence\index{subject}{quantum turbulence!turbulence}, motivated largely by the huge degree of control they offer.  
 Even though the range of lengthscales excited in these systems is much less than in superfluid helium, the direct
 Kolmogorov energy cascade has been predicted to exist and the regimes of turbulence accessible is vast and 
 interesting its own right.  We have distinguished three distinct phases of quantum turbulence - its generation of
 turbulence, its steady state and its decay - and briefly
 reviewed work done in understanding these so far, whilst highlighting fundamental questions about each phase.  
   
 \ack
 We acknowledge A. C. White for useful discussions and for providing us with an image for use in this article.
 This work was supported by the grant EP/I019413/1 from the EPSRC.

%
\bibliographystyle{jfm2}

\end{document}